# First-principles studies of electronic properties in Lithium metasilicate (Li$_2$SiO$_3$)

Nguyen Thi Han[a,b,*], Vo Khuong Dien [a], Ngoc Thanh Thuy Tran [c], Duy Khanh Nguyen [d], Wu-Pei Su [e] and Ming-Fa Lin [a,c,†]

[a]*Department of Physics, National Cheng Kung University, 701 Tainan, Taiwan.*

[b]*Department of Chemistry, Thai Nguyen University of Education, 20 Luong Ngọc Quyen, Quang Trung, Thai Nguyen City, Thai Nguyen Province, Viet Nam.*

[c]*Hierarchical Green Energy Materials (Hi-Gem) Research Center, National Cheng Kung University, Tainan, 70101, Taiwan.*

[d]*Advance Institute of Materials Science, Ton Duc Thang University, 19 Nguyen Huu Tho Street, District 7, Ho Chi Minh City, Viet Nam.*

[e]*Department of Physics and Texas Center for Superconductivity, University of Houston, TX 77204, USA.*

## Abstract

Lithium metasilicate (Li$_2$SiO$_3$) has attracted considerable interest as a promising electrolyte material for potential use in lithium batteries. However, its electronic properties are still not thoroughly understood. In this work, density functional theory calculations were adopted, our calculations find out that Li$_2$SiO$_3$ exhibits unique lattice symmetry (orthorhombic crystal), valence and conduction bands, charge density distribution, and van Hove singularities. Delicate analyses, the critical multi- orbital hybridizations in Li-O and Si-O bonds 2s- (2s, 2p$_x$, 2p$_y$, 2p$_z$) and (3s, 3p$_x$, 3p$_y$, 3p$_z$)- (2s, 2p$_x$, 2p$_y$, 2p$_z$), respectively was identified. In particular, this system shows a huge indirect-gap of 5.077 eV. Therefore, there exist many strong covalent bonds, with obvious anisotropy and non-uniformity. On the other hand, the spin-dependent magnetic configurations are thoroughly absent. The theoretical framework could be generalized to explore the essential properties of cathode and anode materials of oxide compounds.

[*]*han.nguyen.dhsptn@gmail.com*

[†]mflin@mail.ncku.edu.tw

1 **Introduction**

Nowadays, the Li$^+$ based batteries (LIBs) [1,2] have become one of the mainstream systems in the material basic science [3], engineering, and applications, mainly owing to the diverse geometric [4], electronic [5], and transport properties [6]. They are principally composed of the electrolyte [7,8], cathode [9] and anode materials [10]. All in all, each component possesses unusual geometry with a very large primitive unit cell, directly reflecting the complicated chemical bondings. The greatly modulated bond length could be regarded as a very important common characteristic. They should be the critical condition in searching for the optimal match of three kinds of core components. Seemingly, how to achieve the best LIBs with the highest performance, is a unified engineering issue [11]. LIBs are widely utilized in many electronic devices, e.g., cell phones, laptops, iPods, cars/buses [3], and radios [11]. Their main features include high capacity [10], large output voltage [12], long-term stability [11], and friendly chemical environment [13].

The crucial mechanisms of LIBs are characterized by the unique charging and discharging based on the exchange of Li$^+$ ions. Indeed, Li+ ion transports occur at any time during the charging/discharging processes by the path of the cathode (right-hand arrow), electrolyte (right-hand arrow), anode/anode (right-hand arrow), electrolyte (right-hand arrow), and cathode [14]. Specifically, a separator membrane [15] is inserted to avoid the internal short circuit and only accept the smallest Li$^+$ to freely pass the positive and negative electrodes [5]. The two electrodes are linked externally to an electric supply after the initial charging process, in which electron carriers rapidly escape from the cathode and are transported by the external lead to the anode, leading to the creation of a charge current [1]. To keep the electric neutrality, Li$^+$ ions are rapidly transported along the parallel direction internally from the cathode to the anode by the solid/liquid/gluon state electrolyte [16,17]. With this efficient process, the external energy from the electrical supply is stored in the battery in the form of chemical energy. The opposite process, in which the electrons move from the anode to the cathode through the external lead and the Li$^+$ ions move back to the cathode via the specific electrolytes, can provide the electric power and thus do work on electronic devices [5].

In this paper, the main focuses are the geometric symmetries and electronic properties of Lithium metasilicate (Li$_2$SiO$_3$). The first-principles method is available in delicately calculating the total ground state energy, lattice symmetry, distinct Li-O and Si-O bond lengths, the atom-dominated valence and conduction bands, the spatial charge density, and the atom and orbital-projected density of states. The spin-created phenomena, the spin-split band structure across the Fermi level, the net magnetic moment and the spin density distributions, will be thoroughly examined whether they could survive in this emergent material. These physical quantities are very important in achieving the critical multi-orbital hybridizations of two kinds of chemical bonds and the spin-dependent magnetic configuration. Most of the analysis difficulties arise from the very complicated orbital-projected van Hove singularities, which is partially supported by the electronic structure and charge density distribution. The theoretical predictions on the optimal geometry, the occupied electronic states, and the bandgap and whole energy spectrum, could be tested from the high-resolution measurements of X-ray diffraction/low-energy electron diffraction [18], angle resolved photoemission spectroscopy [19], and scanning tunneling microscopy [20], respectively. In addition, the close relationship between the numerical VASP calculations and the tight-binding model is discussed in detail.

## 2 Computational details

The rich and unique geometric structure and electronic properties of $Li_2SiO_3$ compound were carried out using the density functional theory (DFT) [21-24] implemented by the Vienna ab initio simulation package (VASP). The many-body exchange and correlation energies, due to the electron-electron Coulomb interactions, are investigated from the Perdew-Burke-Ernzerh functional (PBE) [25] within the generalized gradient approximation. Moreover, the projector-augmented wave (PAW) [26] pseudopotentials are available in characterizing the significant electron-ion scatterings. Of course, those two critical interactions have no exact solutions in the analytic forms. Therefore, it is difficult to carry out an accurate diagonalization of the many-particle Hamiltonian [18]. Plane waves, with a kinetic energy cutoff of 500 eV, chosen as a basic set [18] would make it more convenient and reliable to solve for the Bloch wave functions and band structures. The first Brillouin zone is sampled by 7x7x7 and 20x20x20 k-point meshes within the Monkhorst - Pack scheme [27] for geometric optimization and electronic energy spectrum, respectively. These wave-vector points should be sufficient in calculating the suitable orbital-projected van Hove singularities, spatial charge distributions, and spin density configurations. Besides, the convergence condition of the ground-state energy is set to be $10^{-5}$ eV between two consecutive simulation steps;

furthermore, the maximum Hellmann-Feynman force acting on each atom is smaller than 0.01 eV under the ionic relaxations. The accurate VASP calculations are very useful in thoroughly exploring certain physical quantities, e.g., the atom-induced valence and conduction bands, the spatial charge densities due to chemical bondings, the atom and orbital-projected density of states, the atom-dependent spin configurations, the spin-split or degenerate states across the Fermi level, and the finite or vanishing magnetic moments.

**3 Results and discussions**

**3.1 Geometric structure**

Lithium metasilicate ($Li_2SiO_3$) material possesses unique lattice symmetry, according to the delicate first-principles calculations on the optimal geometric structures. We have chosen one of the meta-stable systems for studying the rich and unusual phenomena. The calculated lattice constants of Lithium metasilicate are 5.44 Å, 9.46 Å and 4.72 Å for a, b, c direction, respectively and very close to the previous experimental result [28]. Such material, as clearly illustrated in **Figure 2,** directly corresponds to an orthorhombic Cmc21 structure [29] with a primitive unit cell containing 24 atoms (8- Li, 4- Si and 12- O atoms). The various chemical bonds in the unit cell dominate all the fundamental properties. There exist only 32 Li-O and 16 Si-O bonds, and the other kinds of chemical bondings are thoroughly absent. Most importantly, the optimal geometric parameters in **Table 1** clearly illustrate that their bond lengths vary over a range of 1.94 Å - 2.20 Å and 1.61 Å - 1.70 Å and the fluctuation percentages $\Delta b(\%) = |bond_{max} - bond_{min}|/bond_{min}$ are over 13.3 % and 5.6 % for Li-O and Si-O bonds, respectively. Indicating possible structural transformation between two meta-stable systems with distinct geometries. The total ground state energy is -10.0 eV per unit cell, in which the spin-dependent interactions do not make any contributions. Obviously, the chemical/physical/material environments are highly anisotropic and extremely non-uniform, and the other essential properties are expected to behave similarly. The diverse atomic arrangements are easily observed under distinct plane projections, e.g., the geometric structures for (a) (100), (b) (010), (c) (001), (d) (110), (e) (011), (f) (101), (g) (111) **(Figures S1 (a)- S1 (g)).** Li, Si, and O atoms are denoted by the green, purple and red balls, respectively. The above-mentioned real-space lattice gives rise to the orthorhombic first Brillouin zone shown in **Figure 3**, in which the high symmetry points are very useful in characterizing the electronic energy spectra and states. They include Γ (0.00, 0.00, 0.00), X (0.33, 0.33, 0.00), S (0.00, 0.50, 0.00), Y(-0.50, 0.50, 0.00), T (0.50, 0.50, 0.00), Z (0.00, 0.00, 0.50), A (0.33, 0.33, 0.5), and R (0.00, 0.50, 0.00). The high resolution X- ray elastic scatterings [18] and low-energy electron diffractions (LEED) [18] are very suitable for verifying the three dimensional lattice symmetries of LiSiO based compounds, while the opposite is true for the tunneling microscopy (STM) for the nanoscale top views [15] and tunneling electron microscopy (TEM) for the side-view of the structures [17]. Whether the large modulations of the Li-O and Si-O bond lengths can be identified from the measured data are worthy of the systematic investigations. This is closely related to the very complex multi-orbital hybridizations of all the chemical bonds. Similar examinations could be generalized to the other meta- stable or intermediate configurations [30], being very useful in understanding the transformation paths between them [3].

### 3.2 Electronic properties

The Lithium metasilicate presents rich and unique electronic properties. The electronic band structure as clearly illustrated in **Figures 4 (a) - 4 (d)** the strongly depends on the wave-vector. The occupied valence bands are highly asymmetric to the unoccupied conduction bands about the Fermi level $E_F$ =0 (**Figure 4 (a))**, directly reflecting the very complicated multi-orbital hybridizations in Li-O and Si-O bonds. The energy dispersions, which are shown along the high-symmetry point, have strongly anisotropic behaviors. For example, there exist parabolic, oscillatory and partially flat dispersion relations. Furthermore, the subband non-crossing, crossing and anti-crossing phenomena [11] come to exist frequently. Most importantly, the highest occupied and the lowest unoccupied states, respectively, appear at the Z and Γ points (0.00, 0.00, 0.50) and (0.00, 0.00, 0.00). This indicates a very large indirect band gap of $E_g \approx$ 5.077 eV. Such value is only slightly lower than that $E_g \approx$ 5.5 eV of the diamond. It is very important to recognize that the large bandgap ensures the insolating of electrons while providing high ionic conductivity. Which is a crucial feature to avoid the internal circuit. Apparently, the optical threshold absorption frequency, which can be measured from the reflection/absorption/transmission spectroscopies [31,32], should be higher than 5.1 eV [28]. That is to say, various high-resolution optical measurements are very suitable in examining the insulating behaviour [27]. In addition, the spin splitting is present in any energy band, none of Li, Si and O atoms can create spin-up or spin-down configurations. As a result, there is no magnetic moment and the spin density distribution becomes meaningless. In addition to the main features of band structure, the electronic wave functions for valence and conduction states could provide partial information on the chemical bondings. Each band state can be regarded as a linear superposition of different orbitals. Therefore, it can be decomposed into distinct atomic contributions. The different atom dominances, being proportional to the ball radius, are available for understanding the important role played by the chemical bonds in the electronic properties. The green, purple and red balls, respectively, correspond to the Li, Si and O atomic contributions. All three atoms have significant contributions to the whole band structure. This unusual phenomenon may explain the wide range modulations of the chemical bonding strength in Li-O and Si-O bonds (the strongly modulated hopping integrals) [33,34]. However, it might be difficult to observe the Li-contributions (the small green balls) in **Figure 4 (b)** because of the single 2s- orbital. On the other hand, the silicon and oxygen contributions are obvious in the electronic energy spectrum, i.e. the sufficiently large purple and red balls in **Figures 4 (c) and 4 (d)**. Specifically, the oxygen atoms dominate all the valence and conduction states, since they are associated with the entire chemical bonds. On the experimental side, high-resolution ARPES [18] is the only method for examining the wave-vector dependence of occupied electronic states, with the diversified energy dispersion relations in semiconductors/metals [35]. It is also well known to be very successful for various condensed matter systems, such as the $sp^2$ based carbon materials [36,37], three dimensional LiXO ternary compounds [38], superconductors [39], and group-IV, group-V layered systems [40,41]. Very interestingly, a lot of ARPES experiments have verified the valence-band energy spectra of emergent graphene-related systems, being initiated from the K/K' or Γ valleys. For example, there exist the linear Dirac-cone structures in monolayer/twisted bilayer graphenes [42] and monolayer graphene [43], the parabolic/parabolic and linear bands in bilayer/trilayer AB stackings [44], the linear, partially flat and Sombrero-shaped energy spectra in tri-layer ABC stacking [45], the monolayer and bilayer-like behaviors in AB-stacked graphite [46], and the energy gap and parabolic bands of one dimension graphene nanoribbons [47]. Only monolayer graphene and one dimension materials belong to zero or finite-gap semiconductors [48,49]. Apparently, the unusual electronic properties directly reflect the pure and unique systems, being initiated from unique interlayer $2p_z$-$2p_z$ orbital hybridizations in the normal/enlarged/reduced honeycomb lattices [37]. The diverse behaviors are also confirmed by tight-binding model [45]. The ARPES measurements could be utilized to detect the main features of the band structure in lithium metasilicate ($Li_2SiO_3$). The experimental verifications could provide certain information on the multi-orbital hybridizations of chemical bonds. Very interestingly, the multi-orbital hybridizations in Li-O and Si-O chemical bonds could be roughly examined with the spatial charge distributions, as clearly illustrated in **Figures 5 (a) - 5 (g)**. It is well known that the charge densities are very sensitive to the change of bond lengths. Evidently, the Lithium metasilicate has highly modulated chemical bonds (**Figure S1** and **Table 1**), being closed related to the available orbitals in different atoms. Concerning the shortest Li-O bonds 1.943 Å, in **Figure 5 (b)**, each lithium atom only provides a single 2s- orbital, and its effective distribution range is 0.54 Å, as measured from the deep yellow region of the $Li^+$ ion core to the light green one of the outermost orbitals. However, the two 1s orbitals do not take part in the critical orbital hybridizations with the oxygen atoms. The important oxygen orbitals, which correspond to the light green and yellow regions, are approximately predicted to be associated with the ($2p_x$, $2p_y$, $2p_z$) ones. Specifically, the O-2s orbitals are relatively far away from the Li-atom, so they might play a minor role in Li-O chemical bondings resulting in the increase of Li-O length to 2.201 Å. In **Figures 5 (c) and 5 (d)**, the orbital overlap of Li and O atoms declines quickly. As a result, the multi-orbital 2s- (2s, $2p_x$, $2p_y$, $2p_z$) hybridizations of Li-O bonds exhibit the diverse hopping integrals [43]. There are fewer Si-O chemical bonds, but with quite strong bonding strength (**Figures 5 (e) - 5 (g)**) compared with those of Li-O bonds. The effective charge distribution range of Si atom is much higher than that of Li because of the large atomic number. Most important, both silicon and oxygen atoms could be classified into the heavy red and yellow-green regions which, respectively, correspond to the 3s, 2s and ($3p_x$, $3p_y$, $3p_z$), ($2p_x$, $2p_y$, $2p_z$) orbitals. Very interestingly, the chemical bondings of the former are clearly illustrated by the deformed spherical distributions between two atoms. That is to say, the Si-O chemical bonds consist of the four- orbital hybridizations due to (3s, $3p_x$, $3p_y$, $3p_z$) - (2s, $2p_x$, $2p_y$, $2p_z$). The chemical bonding strength declines with the increase of bond length, as indicated by the reduced charge density between Si and O atoms. In addition, the charge density plots in **Figures 5 (e)-5 (g)** show a dense mixture of charge density between the Si and O atoms, indicating a distinct covalent bonding between Si and O atoms. In contrast, as shown in **Figures 5 (b) - 5 (d)**, there is no obvious electron density accumulation in the region between Li and O. We conclude that the bonding between the Li and O atoms is ionic in nature.

In the current work, the atom and orbital-decomposed van Hove singularities are very useful in directly resolving the critical multi - orbital hybridizations in the Li-O and Si-O bonds, as shown in **Figures 6(a) - 6 (d**). Three kinds of atoms have the significant contributions within the whole energy spectrum between -8.0 eV and 8 eV, reflected in the large modulation of bond lengths **(Table 1)**. For a wide-gap three dimensional $Li_2SiO_3$ compound, the density of states per unit cell is vanishingly small within a large energy range of $E_g \approx 5.077$ eV centered about the Fermi level (the black curve) in **Figure 6 (a)**. The outstanding insulation property may on pose a problem for further STS measurements (later discussions) [50]. The DOS is dominated by the valence states (E<0), but not the conduction ones (E>0). That is, the hole and electron energy spectra are highly asymmetric about the Fermi level. The van Hove singularities mainly arise from the valence and conduction band-edge states along the high-symmetry-point paths, corresponding to the local minimum, maximum, saddle points in the reciprocal space. At least, ten and six special features are, respectively, revealed in the valence and conduction energy spectra. As a result of the frequent crossings and anti-crossings, each feature reflects the combined phenomenon of certain neighboring band-edge states under finite broadening effects. This indicates the experimental difficulty to examine the number of the energy subbands and their band-edge states.

Seemingly, the lithium and oxygen atoms (the green and red curves), respectively, make the weakest and strongest contributions to the total density of states. Such result is principally determined by the number of available atom orbitals in a primitive unit cell. Generally speaking, the important contributions, which arise from the various orbitals of different atoms, cover (I) Li-2s orbitals (the purple curve) in **Figure 6 (b)**, (II) Si-($3s, 3p_x, 3p_y, 3p_z$) orbitals (the red, blue, green and purple curves) in **Figure 6 (c)**, and (III) O-($2s, 2p_x, 2p_y, 2p_z$) orbitals (the red, blue, green and purple curves) in **Figure 6(d)**. All the orbital contributions are merged together, especially for the number, energies, intensities and forms of van Hove singularities. The main reason is that the distinct chemical bondings are associated with one another through the Li-O and Si-O bonds. Consequently, there exist the multi-orbital hybridizations of 2s- ($2s, 2p_x, 2p_y, 2p_z$) and ($3s, 3p_x, 3p_y, 3p_z$) - ($2s, 2p_x, 2p_y, 2p_z$). The enlarged density of states for Li- 2s orbitals can provide the 5 clear van Hove singularities and illustrates its weak, but significant feature in **Figure 6(b).** The observable contributions are obviously revealed in the half-occupied four orbitals of silicon atoms (**Figure 6(c))**, in which their differences only reflect the strong anisotropy in a primitive unit cell (**Figure 2**).

Most importantly, only three orbitals of O- ($2p_x$, $2p_y$, $2p_z$) make comparable and dominating contributions (**Figure 6 (d)**), while the 2s ones are much smaller than the others in the opposite region. This unusual behavior is consistent with the (3s, $3p_x$, $3p_y$, $3p_z$)- (2s, $2p_x$, $2p_y$, $2p_z$) and 2s- ($2p_x$, $2p_y$, $2p_z$) orbital hybridizations in the Si-O and Li-O bonds, respectively.

Roughly speaking, when a quantum tunneling current (a very small I) is injected from a nanoscaled probe onto a sample surface by applying a finite potential difference the measured differential conductance dI/dV, could be regarded as the density of states. Very interestingly, the up-to-date STS has been successful in identifying the dimension-enriched van Hove singularities of graphene-related systems with the $sp^2$ chemical bondings. Similar STS experiments, for lithium metasilicate ($Li_2SiO_3$) materials could verify a very large energy gap of $E_g \approx 5.077$ eV, six/three asymmetric/symmetric peaks and broadening shoulders in valence/conduction spectrum, the high asymmetry about electron and hole states and their distinct distribution widths. The experimental measurements, together with the theoretical predictions on van Hove singularities would allow one to identify the complicated band structure and thus multi-orbital hybridizations in Li-O and Si-O bonds. In the current $Li^+$ based batteries [51], Lithium metasilicate ($Li_2SiO_3$) material could serve as the solid-state electrolytes [52]. During the charging process, a plenty of lithium ions are rapidly transferred from the cathode (e.g., Li/Fe/Co/NiO materials) [53,54,55], into the electrolyte and then anode. The opposite is true for the discharging process. Seemingly, the $Li^+$ flowing will generate the dramatic transformation of any materials. Our calculation might help enhance the understanding of the Li motion.

### 4 Concluding remarks

The rich and unique properties of the emergent lithium metasilicate ($Li_2SiO_3$) are thoroughly investigated from the first-principles calculations. The critical multi-orbital hybridizations, which survive in Li-O and Si-O bonds, are accurately identified from the atom-dominated band structure, the charge density distributions in the greatly modulated chemical bonds, and the atom- and orbital- projected van Hove singularities. The theoretical framework could be further developed for other electrolytes, anode and cathode materials of $Li^+$ based batteries [52], e.g., the important differences among the various LiXO related compounds [38], and the diverse phenomena driven by the various components. Very interestingly, the highly anisotropic and non-uniform environments need to be included in the phenomenological models. The solid states electrolyte material of $Li_2SiO_3$ with 24 atoms in a primitive unit cell, is an orthorhombic structure. There are 32 Li-O and 16 Si-O chemical bonds, in which each atom has four neighboring other atoms. Most importantly, the very strong orbital hybridizations create a very wide indirect gap of $E_g \approx 5.077$ eV, being close to the largest one in diamond $\approx 5.5$ eV. Furthermore, there exist very large band width (-8.0 eV, $E^{c,v}$, 8.0 eV ), strong/various energy dispersion, high anisotropy, and frequently non-crossing/crossing/anti-crossing behaviors. They all contribute to the broadening of asymmetric/symmetric peaks and shoulders and play a critical role in examining the multi-orbital hybridizations of Li-O and Si-O bonds, 2s- ($2s, 2p_x, 2p_y, 2p_z$) and ($3s, 3p_x, 3p_y, 3p_z$)- ($2s, 2p_x, 2p_y, 2p_z$). The diverse covalent bondings are partially supported by the atom-dominated band structure and charge density distributions in modulated chemical bonds. The theoretical predictions on the optimal geometry, wave-vector-dependent valence bands, and valence and conduction density of states, could be verified by X-ray elastic scatterings, LEED [18] and ARPES [56]

**Acknowledgement**

This work was financially supported by the Hierarchical Green-Energy Materials (Hi-GEM) Research Center, from The Featured Areas Research Center Program within the framework of the Higher Education Sprout Project by the Ministry of Education (MOE) and the Ministry of Science and Technology (MOST 108-3017-F-006 -003) in Taiwan.

Table 1 The bond lengths in Li$_2$SiO$_3$ for the total number of 32 Li-O and 16 Si-O bonds

| Atom-Atom | No. of bonds | Bond length (Å) | |
|---|---|---|---|
| | | Experiment [28] | This work |
| Li-O | 32 | 1.932-2.176 | 1.943-2.201 |
| Si-O | 16 | 1.591-1.680 | 1.612-1.701 |

**Figure captions**

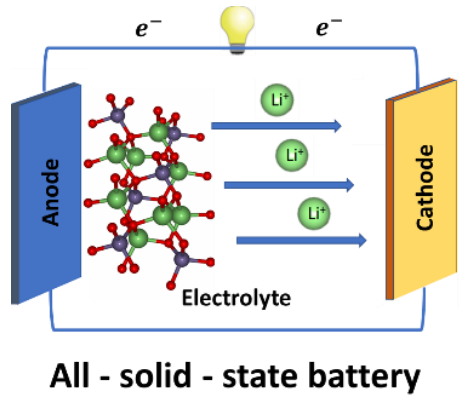

**Figure 1** All -solid-state Li⁺ based battery with the electrolyte of three dimesion ternary Li₂SiO₃

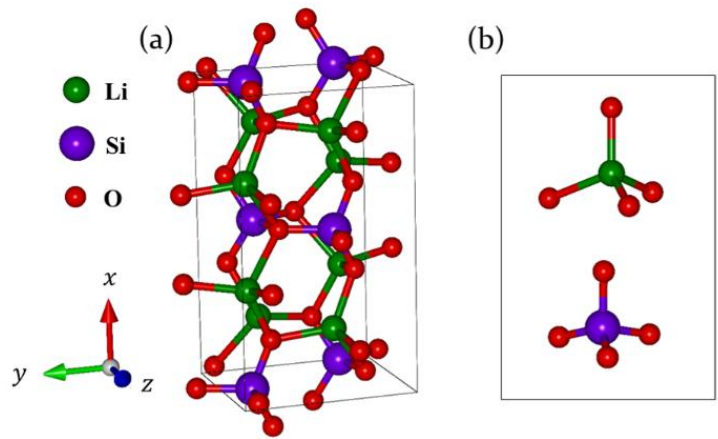

**Figure 2** (a)The optimal lattice structure of $Li_2SiO_3$ under the orthorhombic symmetry with 24 atoms in a unit cell of lattice constants a = 5.440 Å, b = 9.467 Å and c = 4.719 Å (b) Two kinds of chemical bondings: Li-O and Si-O.

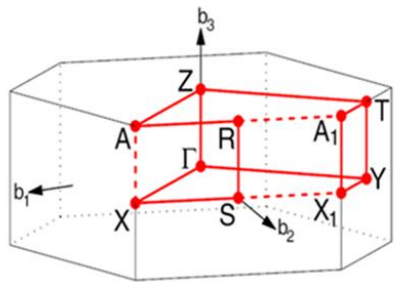

**Figure 3** The first Brillouin zone with the high-symmetry points within the three orthogonal axes.

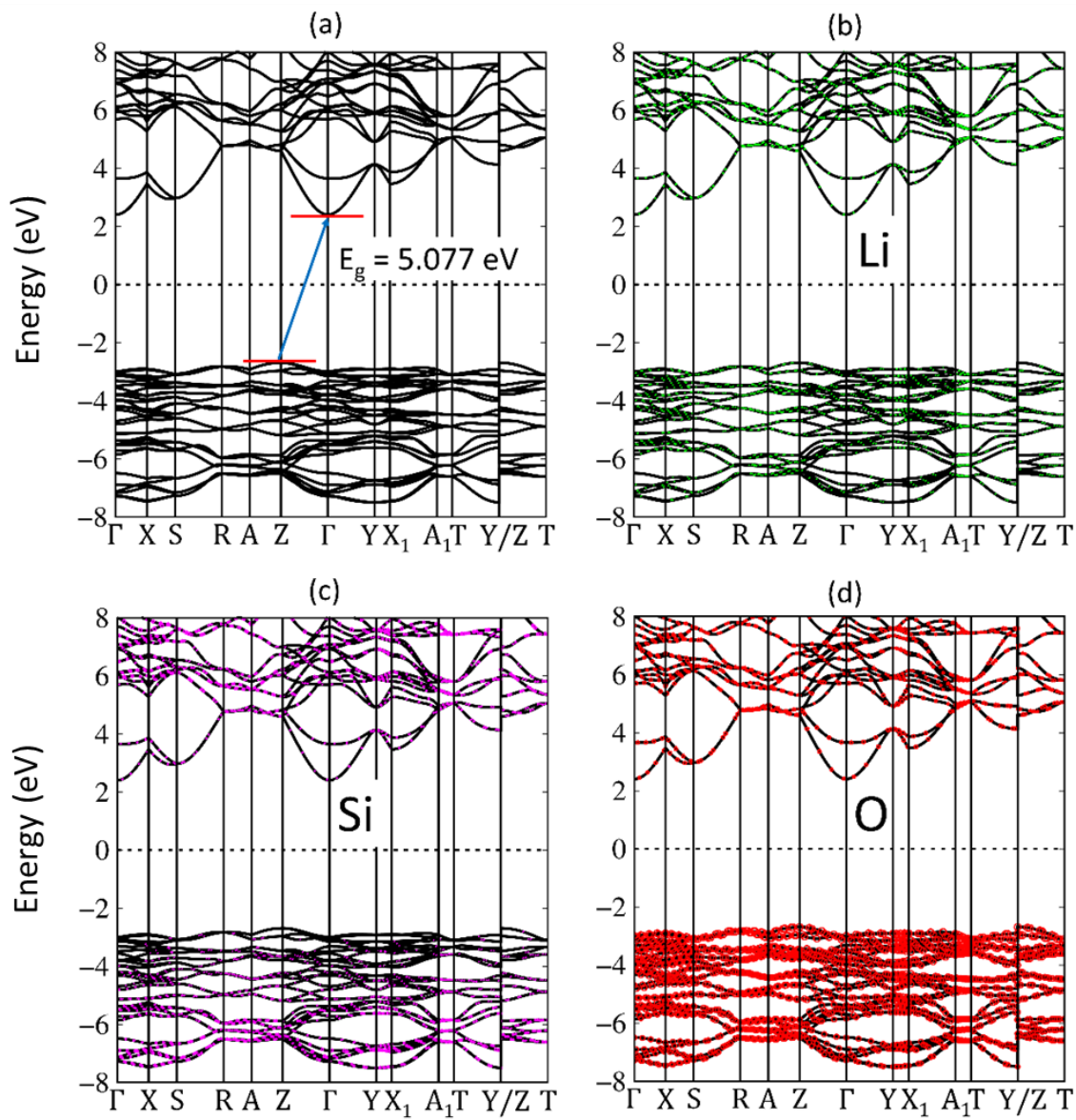

**Figure 4** (a) The significant valence and conduction band of $Li_2SiO_3$ along the high-symmetry points within the first Brillouin zone for the energy range (-8.0 eV, $E^{c,v}$, 8.0 eV) with the specific (b) lithium, (c) silicon and (d) oxygen dominances (green, purple, and red balls), respectively.

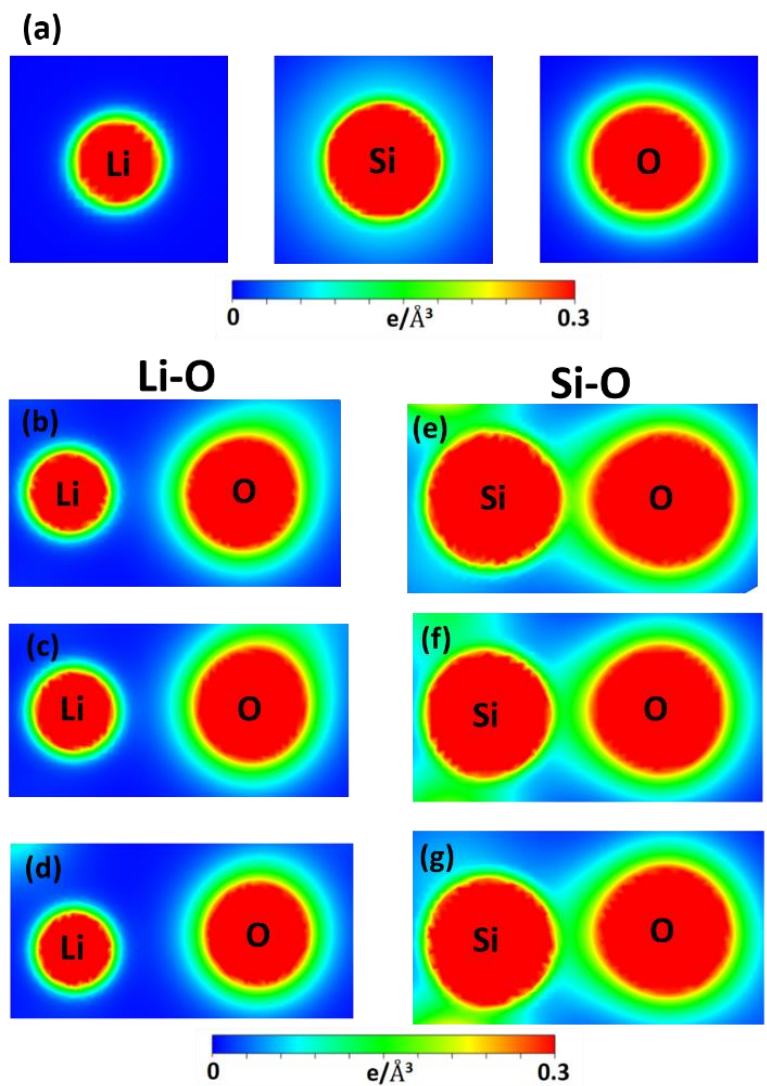

**Figure 5** The spatial charge density distribution for (a) the isolated (Li, Si, O) atoms. The shortest/medium/longest (b)/(c)/(d) Li-O and (e)/(f)/(g) Si-O bonds

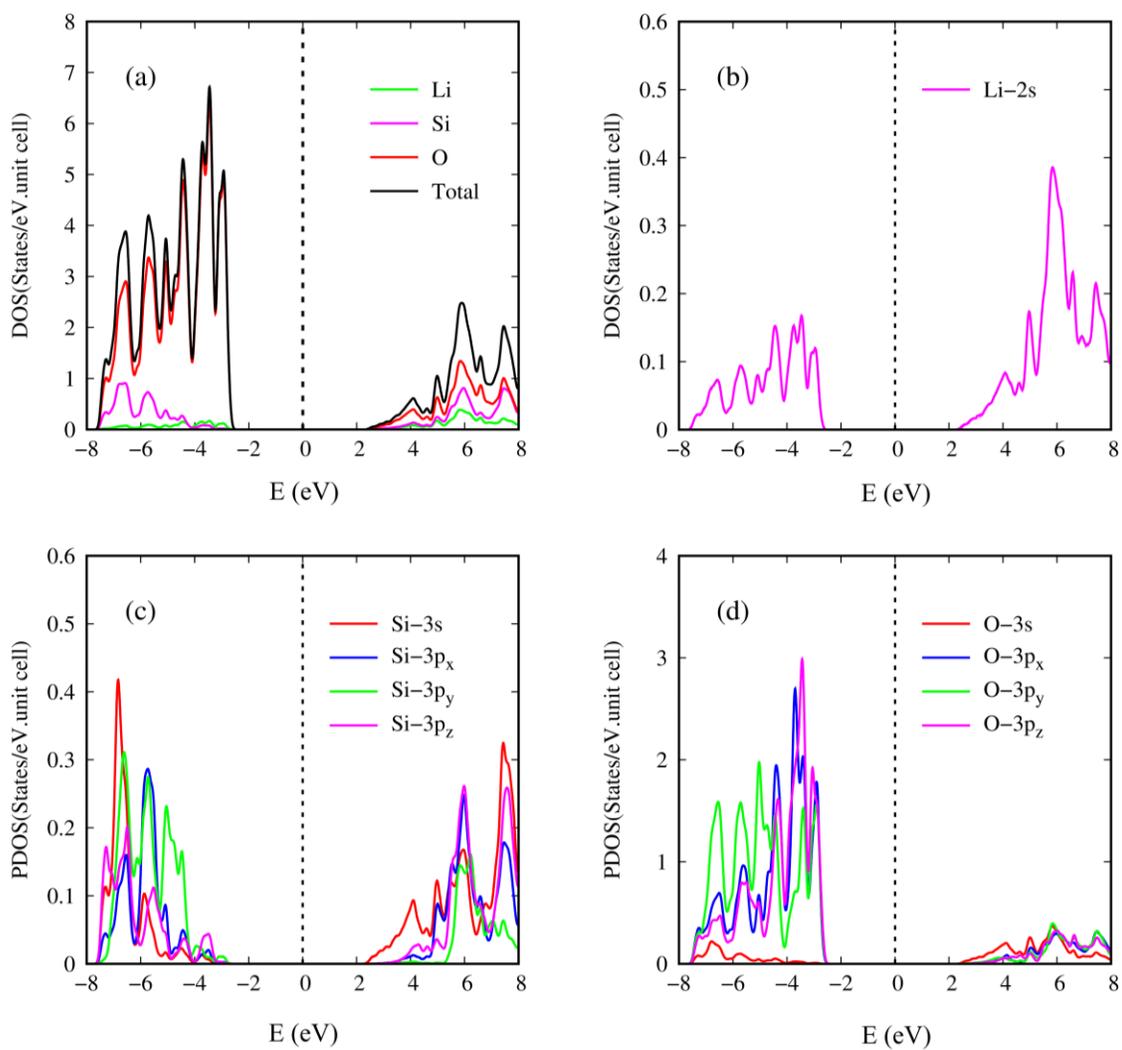

**Figure 6** The atom and orbital-projected density of state: those coming from (a) Li, Si, and O atoms, (b) Li-2s orbital, (c) Si-(3s, 3p$_x$, 3p$_y$, 3p$_z$) orbitals, and (d) O- (2s, 2p$_x$, 2p$_y$ and 2p$_z$) orbitals

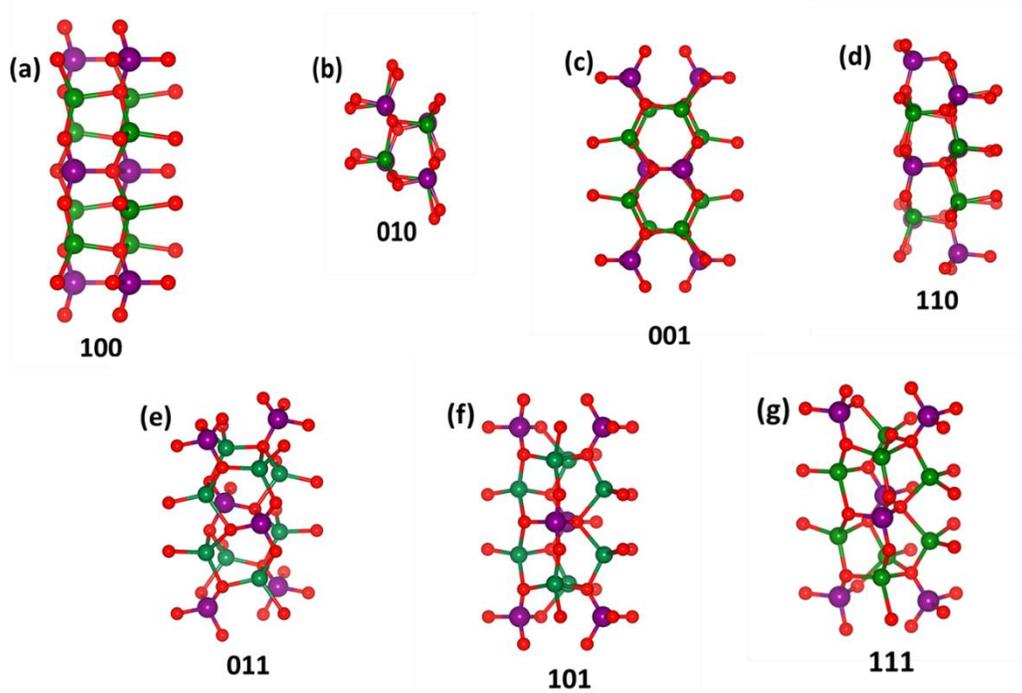

**Figure S1(a)-S1(g)** Projections of $Li_2SiO_3$ onto distinct planes: (a) (100), (b) (010), (c) (001), (d) (110) (e) (011), (f) (101), (g) (111) where the Li, Si and O atoms correspond to the green, purple and red balls, respectively.